\documentclass{article}
\usepackage{listings}
\usepackage{authblk}
\usepackage{amsmath}
\usepackage[margin=1in, top=0.5in]{geometry}
\usepackage{enumitem}
\usepackage[]{graphicx}
\usepackage{caption}
\usepackage{subcaption}
\usepackage{url}  
\usepackage[colorlinks = true]{hyperref} 
\usepackage{algpseudocode,algorithm,algorithmicx}

\algrenewcommand\algorithmicrequire{\textbf{Precondition:}}
\algrenewcommand\algorithmicensure{\textbf{Postcondition:}}
\usepackage{bm}
\usepackage{amsfonts}   
\usepackage{amsmath} 
\usepackage{amsthm} 
\usepackage[superscript,biblabel]{cite}

\title{A general framework combining individual cancer models to project downstaging due to multi-cancer early detection testing}
\author{Jane M. Lange$^1$ \and Kemal C. Gogebakan$^2$ \and Roman Gulati$^2$ \and Ruth Etzioni$^{2,3}$}

\date{%
    $^1$Oregon Health and Science University\\%
    $^2$Fred Hutchinson Cancer Research Center\\
    $^3$University of Washington, Department of Health Services\\[2ex]%
    June 30, 2023
}
\begin{document}
\maketitle
\begin{abstract}
Multi-cancer early detection (MCED) tests offer to screen for multiple types of cancer with a single blood sample. Despite their promising diagnostic performance, evidence regarding their population benefit is not yet available. Expecting that benefit will derive from detecting cancer before it progresses to an advanced stage, we develop a general two-stage model to project the reduction in advanced-stage diagnoses given stage-specific test sensitivities and testing ages. The model can be estimated using cancer registry data and assumptions about overall and advanced-stage preclinical sojourn times. We first estimate the model for lung cancer and validate it using the stage shift observed in the National Lung Screening Trial. We then estimate the model for liver, pancreas, and bladder cancer, which have no recommended screening tests, and we project stage shifts under a shared MCED testing protocol. Our framework transparently integrates available data and working hypotheses to project reductions in advanced-stage diagnoses due to MCED testing.
\end{abstract}

\section{Introduction}

Multi-cancer early detection (MCED) tests offer to screen for multiple types of cancer with a single blood draw. Multiple studies have reported promising diagnostic performance in clinically diagnosed patients \cite{Cohen2018} and in prospectively screened cohorts \cite{Lennon2020,Nadauld2021}, yet no clinical trials of MCED testing have been completed. The US National Cancer Institute established the Cancer Screening Research Network to conduct trials of MCED tests on cancer mortality, but even preliminary results will take many years.

The reduction in advanced-stage diagnoses has been proposed as an early indicator of mortality benefit. Although this so-called ``stage shift'' has not been established as a bona fide surrogate, it is widely understood to be a necessary condition for mortality benefit. Recent work has shown a positive but imperfect correlation between these outcomes in selected cancers with existing screening tests.\cite{Owens2022} Importantly, a stage shift is also associated with reduced morbidity and lower healthcare costs. An ongoing trial to reduce advanced-stage diagnoses has completed enrollment in the UK\cite{Neal2022}, but results from this study will be limited to the particular MCED product being used and under the particular screening protocol in this population.

Modeling has become an established tool to address gaps in the evidence base when screening trials are lacking or inadequate and for extrapolating outcomes to longer time frames and under different screening protocols.\cite{Habbema2014} A central challenge with modeling MCED testing is that many target cancers have no established screening tests, and consequently little is known about the preclinical latency period within early and advanced stages. Under working hypotheses about the mean overall and advanced-stage latency periods and data on age- and stage-specific incidence rates, however, the parameters of a cancer natural history model can be estimated.

In this study, we develop a two-stage progressive cancer model that can be estimated using age- and stage-specific incidence data from the Surveillance, Epidemiology, and End Results (SEER) registry and hypotheses about preclinical latency periods. Using the estimated model, we project stage shifts from a screening test given sensitivities to detect preclinical tumors and specified testing ages. We illustrate the model estimation procedure for lung cancer, validate the model by comparing results to the National Lung Screening Trial (NLST),\cite{Team2011} and demonstrate how model estimates for single cancers can be aggregated to project stage shifts under a shared MCED testing protocol. Additional analyses evaluate how the stage shift depends on hypothesized sojourn times, stage-specific test sensitivities, and testing intervals. Our framework represents a pragmatic test bed for integrating available data and working hypotheses to translate general MCED test products and protocols to reductions in advanced-stage diagnoses.
 
\section{Methods}
\subsection{General model structure}

\begin{figure}[htbp]
    \centering
    \includegraphics[width=\textwidth]{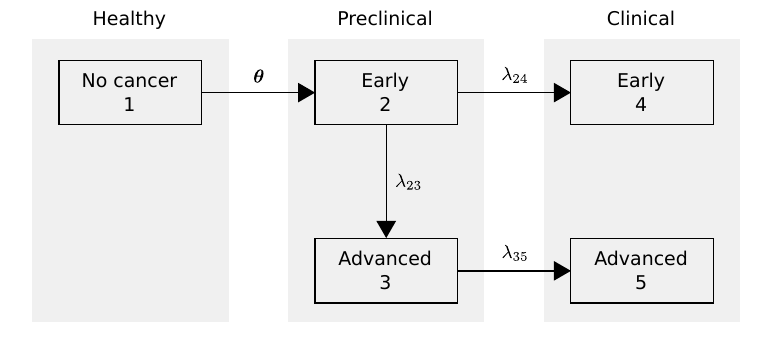}
    \caption{General two-stage progressive natural history model}
    \label{fig:natural historymain}
\end{figure}

Classical models of cancer natural history include healthy, preclinical and clinical states.\cite{Walter1983,Zelen1969,Duffy1995,Shen2001} To project reductions in advanced-stage diagnoses, however, in this study we consider the simplest natural history model that distinguishes early and advanced stage cancers in the preclinical and clinical states (Figure~\ref{fig:natural historymain}).\cite{Pinsky2004}.

To balance model flexibility and computational efficiency, we specify a flexible hypoexponential distribution for the transition rate from healthy to early-stage preclinical cancer, which is governed by the parameter vector $\boldsymbol{\theta}$ (Supplement \ref{sup:model}). This formulation of the rate of onset of a preclinical tumor follows the Armitage-Doll multistage model for carcinogenesis.\cite{Armitage1954} We also assume constant rates of preclinical stage progression (i.e., transitioning from state 2 to 3) and clinical diagnosis (i.e., transitioning from 2 to 4 or from 3 to 5), which we represent as $\lambda_{23}$, $\lambda_{24}$, and $\lambda_{35}$.

The model can be estimated using age- and stage-specific incidence rates from the SEER cancer registry. However, incidence data are not enough to uniquely estimate the model parameters. To fit the model, we assume working hypotheses for the overall mean preclinical sojourn time (OMST) and the mean late-stage sojourn time (LMST) are available as inputs. In the general natural history model, these can be expressed as:
\[
\begin{aligned}
\text{LMST}&=\frac{1}{\lambda_{35}}\\
\text{OMST}&=\frac{1}{\lambda_{24}+\lambda_{23}}+\frac{\lambda_{23}}{\lambda_{23}+\lambda_{24}}\frac{1}{\lambda_{35}}.
\end{aligned}
\]

\noindent
Given incidence data and hypothesized values of LMST and OMST, all unknown disease progression parameters can be estimated (Supplement \ref{sup:identifiability}). Further, the early mean sojourn time (EMST), which represents the window of opportunity for early detection, is completely determined:
\[
\text{EMST}=\frac{1}{\lambda_{23}+\lambda_{24}}.
\]
  
\subsection{Fitting the model}
 
The data used to fit the natural history model consist of counts of early- and advanced-stage diagnoses tabulated by five-year age groups and corresponding population counts. The number of early or advanced-stage diagnoses in each age group is assumed to follow a Poisson distribution with a mean equal to the person-years at risk multiplied by the hazard of early or advanced-stage clinical diagnosis at the midpoint in that age interval. The observed hazards of early- and advanced-stage clinical diagnosis are given by the age- and stage-specific incidence rates. The modeled hazards correspond to derived rates of competing events of early and late clinical states (Supplement \ref{sup:hazard},\ref{sup:latenthazard})
The model-fitting procedure identifies the set of transition rate parameters that provide the best fit between the observed and modeled incidence of early- and advanced-stage clinical diagnosis by maximizing the associated Poisson likelihood (Supplement \ref{sup:likelihood}).

\subsection{Projecting the stage shift}

To project the stage shift from a screening test, we simulate a screening trial with screening in the intervention arm at ages $a_1,\ldots,a_n$, no screening in the control arm, and last follow-up at age $a_{n+1}$. Test sensitivities for early- and advanced-stage preclinical disease are denoted $B_e$ and $B_\ell$. Preclinical disease can be detected early by a screening test or clinically diagnosed between screens or after screening has ended. Supplement \ref{sup:stageshift} provides expressions for the probability of advanced-stage cancer in each arm at each screen or interval. The projected stage shift is the relative reduction in advanced-stage diagnoses between the screen and control arms at the end of the trial.

\subsection{Estimating the model with lung cancer data}
 
 Our lung cancer model defined early vs advanced stages as American Joint Committee on Cancer 7th Edition (AJCC7) categories I--II vs III--IV. To examine sensitivity to hypothesized sojourn times, we considered OMST=2 or 4 years and LMST=0.5 or 1 year. For a wide range of choices for the sojourn times, the dimension of the hypoexponential onset distribution was found to be highly robust based on a visual assessment of the implied concordance with incidence data. To reflect the increased risk of NLST participants, which had at least 30 pack-years of smoking history, we first increased the total incidence counts by a factor of 3 based on a recent analysis\cite{Gogebakan2023}. Given the external inputs, risk-inflated age- and stage-specific incidence data from SEER for the period 2002--2009 (in which lung cancer screening was negligible), and the dimension of the onset distribution, remaining model parameters were estimated using maximum likelihood.

\subsection{Validating the model fit to lung cancer data}

As a proof-of-principle validation of our model fit to lung cancer data, we compared the stage shift projected using our model with the stage shift observed in the NLST.   We then based OMST and LMST hypothesized values via model parameters from the MISCAN-Lung model\cite{TenHaaf2015}, averaging across sex and histologic type (Supplement \ref{sup:weightedparam}). Finally, we re-estimated the endogenous model parameters.

To replicate the NLST design, we simulated a cohort of undiagnosed individuals age 61 years, which was the mean age at the first screen. Individuals were tracked over a 3-year active screening phase plus 4.5 years of follow-up. Because the NLST compared 3 annual screens using low-dose computed tomography (LDCT) vs x-ray radiography during the active screening phase, we projected outcomes under LDCT and x-ray radiography and then compared advanced cases in the two arms.

\subsection{Sensitivity to early stage sojourn times, test sensitivity, and follow-up}

We projected stage shifts under varying assumptions about the EMST, test sensitivity, and follow-up after the last screen using our increased-risk lung cancer model. Specifically, we varied OMSTs from 1 to 7 years (keeping LMST=0.5 years), determined the resulting EMST, varied the early-stage test sensitivity from 0.1 to 1.0 (keeping late-stage sensitivity=.9), and varied the follow-up after the last screen from 1 to 4.5 years, recalculating the stage shift (for screening versus no-screening) for each setting.

\subsection{Combining models of multiple cancers}

To demonstrate our framework for MCED testing of cancers with no existing screening programs, we fit models for liver, pancreas, and bladder cancer using SEER incidence data for the period 2011--2015 under hypothesized sojourn times that reflect slower (OMST=5 years and LMST=1 year) and faster (OMST=2 years and LMST=0.5 years) progression. Early vs advanced stage was defined as AJCC7 I--II vs III--IV for liver and bladder cancer and I vs II--IV for pancreatic cancer because only stage I is typically resectable.\cite{amin2018ajcc} We then projected stage shifts under 3 annual or biennial screening exams starting at age 60 years. We assumed an early-stage test sensitivity of 30\% or 70\% and an advanced-stage sensitivity of 90\%.

\section{Results}

\subsection{Fitted lung cancer models}

\begin{figure}[ht]
    \centering
    \includegraphics[width=\textwidth]{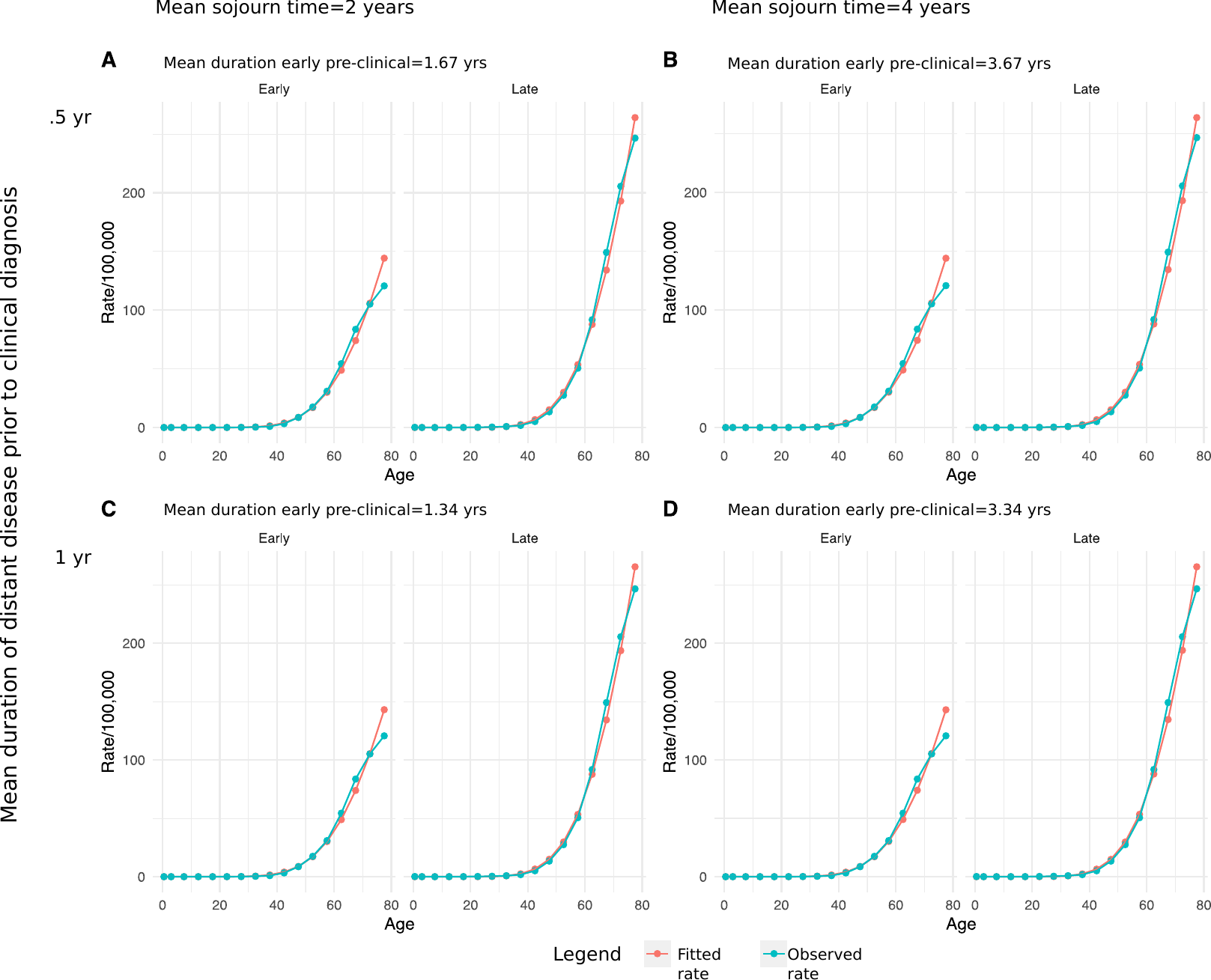}
    \caption{Early- and late-stage lung cancer incidence rates by age observed in the Surveillance, Epidemiology, and End Results program in 2002--2009 and corresponding projections from models fit under assumed mean overall sojourn times of 2 or 4 years and mean late-stage sojourn times of 0.5 or 1 years.}
    \label{fig:incidence}
\end{figure}

Figure~\ref{fig:incidence} displays fit of the lung cancer model to SEER incidence data under varying assumptions about the OMST and LMST. Each model fit used a hypoexponential onset distribution with 11 parameters, which were re-estimated with the endogeneous transition rate parameters. Fits to observed incidence rates are similar across settings, although the implied EMSTs and the cumulative incidence of preclinical onset in the absence of other-cause mortality by age are especially sensitive to assumed OMST (Table~\ref{tab1}). 

\begin{table}\caption{Age-specific cumulative incidence per 100,000 individuals of preclinical onset and clinical diagnosis projected from lung cancer models estimated with SEER incidence data and variable assumptions about the overall mean sojourn time (OMST) and advanced-stage mean sojourn time (LMST). For each setting, the implied early-stage mean sojourn time (EMST) is also shown.}\label{tab1}
\centering
\begin{tabular}{ccccccccccccc}
\hline
\multicolumn{3}{c}{Sojourn times} & & \multicolumn{4}{c}{Preclinical onset by age} & & \multicolumn{4}{c}{Clinical diagnosis by age} \\
\cline{1-3}\cline{5-8}\cline{10-13}
\multicolumn{1}{c}{OMST} & \multicolumn{1}{c}{LMST} & \multicolumn{1}{c}{EMST} & & 50 & 60 & 70 & 80 & & 50 & 60 & 70 & 80 \\
\hline
2 & 0.5 & 1.67 & & 273 & 1077 & 3058 & 6846 & & 203 & 858 & 2560 & 5952 \\
2 & 1.0 & 1.34 & & 272 & 1075 & 3059 & 6859 & & 202 & 855 & 2558 & 5958 \\
4 & 0.5 & 3.67 & & 342 & 1298 & 3561 & 7740 & & 202 & 857 & 2562 & 5952 \\
4 & 1.0 & 3.34 & & 345 & 1306 & 3576 & 7751 & & 201 & 857 & 2563 & 5950 \\
\hline
\end{tabular}
\end{table}

\subsection{Validation of a fitted lung cancer model}

Based on weighted averages of published estimates test sensitivity by sex, stage, and histology, we calculated that LDCT has early- and advanced-stage sensitivities of 35\% and 81\% and x-ray radiography has corresponding sensitivities of 13\% and 56\% (Supplement 3). Based on weighted averages of the MISCAN-Lung model parameters, we assumed hypothesized OMST=4.0 years and LMST=1.4 years. Fitting the natural history model implied EMST=3.1 years, which is close to the MISCAN-Lung model estimate of EMST=3.0 years. For this model fit in a simulation of the NLST, the projected stage shift due to LDCT compared to x-ray radiography was 17\%, moderately lower than the observed stage shift of 21\%.  Compared to no screening, LDCT was associated with a stage shift of 25\%.

\subsection{Sensitivity analyses}

\begin{figure}[ht]
    \centering
    \includegraphics[width=\textwidth]{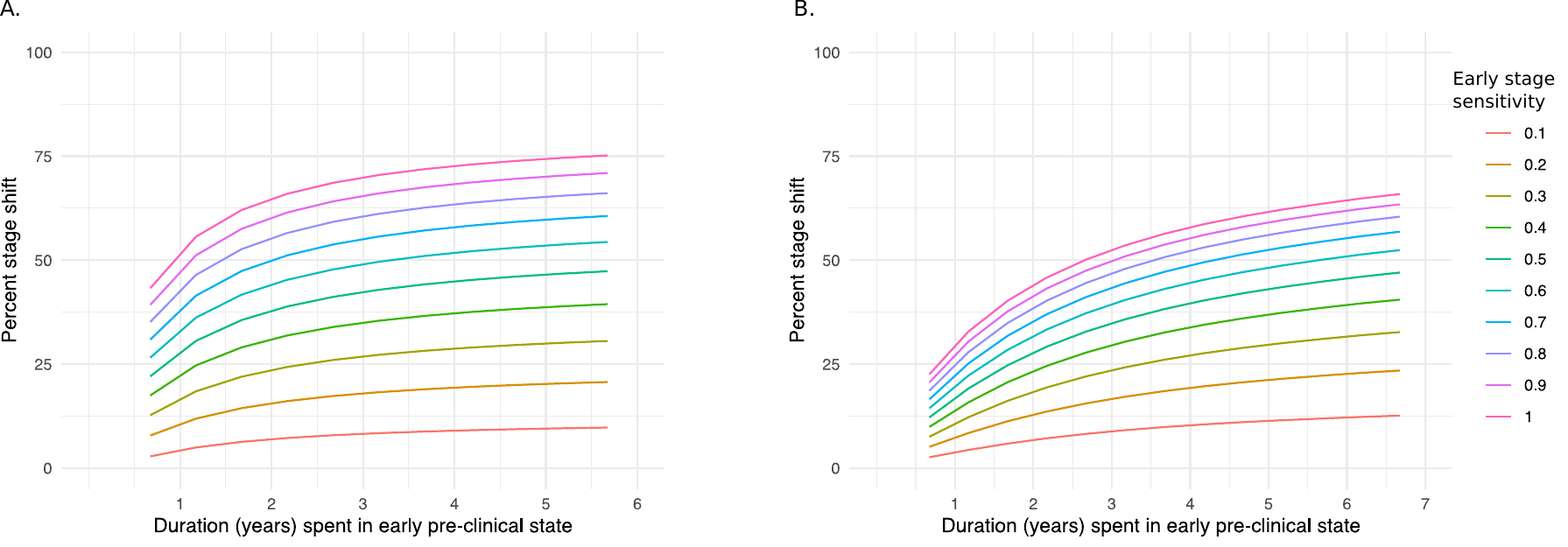}
    \caption{Projected stage shifts in simulated lung cancer trial with three annual screens at ages 60--62 years of age under varying early-stage mean sojourn times (x-axis) and test sensitivities (line colors) under (A) 1 year or (B) 4.5 years of follow-up after the last screen.}
    \label{fig:stageshift}
\end{figure}

Figure~\ref{fig:stageshift} illustrates the impact of varying assumptions about the EMST and test sensitivity on the stage shift in the lung cancer models in  hypothetical screening trials with 1 or 4.5 years years of follow-up after the last screen. A longer EMST, higher test sensitivity, and shorter follow-up after the last screen produce the largest stage shifts. With a perfectly sensitive test, as the EMST varies from X (when the OMST=1 years) to X (when the OMST=7 years), the stage shift ranges from 43\% to 76\% with 1 year of follow-up and from 23\% to 66\% with 4.5 years follow-up after the last screen.

\subsection{Combined testing for multiple cancers}

Table~\ref{tab::othersites} shows projected stage shifts under MCED testing for liver, pancreas, and bladder cancer under varying hypotheses about the OMST and LMST, early-stage test sensitivities, and testing frequencies. A higher early-stage test sensitivity is projected to yield substantially larger stage shifts, and more frequent screening is especially important when mean sojourn times are shorter.

\begin{table}
\centering
\caption{Projected stage shifts under multi-cancer early detection testing for three organ sites under varying hypothesized sojourn times, test sensitivities, and testing frequencies. In each setting, individuals start screening at age 60 and receive 3 annual or 3 biennial screens with 1 year of follow-up after the last screen.}\label{tab::othersites}  
\begin{tabular}{cllcc}
\hline
Early-stage     & Test      & Organ & \multicolumn{2}{c}{Projected stage shift, \%} \\
\cline{4-5}
sensitivity, \% & frequency & site  & OMST=5, LMST=1 & OMST=2, LMST=0.5  \\
\hline
30 & Annual   & Liver    & 29.9 & 24.3 \\
   &          & Bladder  & 27.3 & 24.6 \\
   &          & Pancreas & 26.4 & 22.3 \\
   & Biennial & Liver    & 26.3 & 16.0 \\
   &          & Bladder  & 25.4 & 15.7 \\
   &          & Pancreas & 24.3 & 13.7 \\
70 & Annual   & Liver    & 55.0 & 48.3 \\
   &          & Bladder  & 53.8 & 48.6 \\
   &          & Pancreas & 52.3 & 45.0 \\
   & Biennial & Liver    & 51.3 & 24.3 \\
   &          & Bladder  & 49.7 & 33.4 \\
   &          & Pancreas & 48.0 & 29.8 \\
\hline
\end{tabular}
\end{table}

\section{Discussion}

Assessing the benefit of novel cancer screening tests is challenging. Cancer screening trials are lengthy, costly, and can only evaluate a small number of test products and strategies. With many MCED products in the pipeline, each targeting different sets of cancers and operating with different performance characteristics, flexible methods for synthesizing evidence and bracketing plausible outcomes are needed. Recognizing that screening benefit will likely arise by shifting cancers that would have been diagnosed in an advanced stage to an early stage that is more amenable to curative treatment, we develop a general two-stage natural history model that can be estimated using available incidence data and working hypotheses about latency periods. Estimated models for a set of targeted cancers can be combined to project stage shifts expected under a common MCED testing strategy.

Our general model builds on classical models of cancer natural history that are typically calibrated using incidence data from screened cohorts.\cite{Pinsky2004} Since one of the most exciting possibilities of MCED tests is that they may be able to detect cancers for which data from screened cohorts are not available, we consider model estimation using cancer registry data, together with minimal working assumptions to ensure that parameters are estimable. These assumptions can be varied, and the model re-estimated, to produce a range of projected stage shifts under a specified screening program.

Our framework combines models assuming independence of risks of preclinical onset and progression across cancers. It also represents test sensitivities for individual cancers, which is simplification of front-line and confirmation testing protocols that are likely to be used in practice. Accounting for correlated natural histories (e.g. due to genetic syndromes or risk factors like smoking) or more detailed algorithms for resolving cancer signals would require extending our framework to incorporate these complexities.

The aim of our model is similar to the ``interception'' model of Hubbell et al.\cite{Hubbell2021, Sasieni2023}, which also projects stage shifts based on simulated MCED testing. The main difference with our approach is that Hubbell et al. quantified stage shift by considering the consequences of shifting diagnoses earlier. This approach examines an interesting question: ``What if advanced-stage cancers could have been diagnosed earlier?'' In contrast, our framework superimposes a prospective screening strategy on life histories and isolates  relevant policy levers, such as testing ages and intervals, as explicit exogenous control variables. This approach examines a different question: ``What if this MCED product is used in this screening program in this population?''

Our framework has several applications. First, it can generate realistic expectations about the reduction in advanced-stage diagnoses under single-cancer or MCED testing. Few studies have estimated the sensitivity of MCED tests in prospectively screened cohorts; those that have done so have suggested that sensitivity is likely to be considerably more modest than that estimated from retrospective case-control studies. For example, the sensitivity of the Detect-A MCED test was 27\% in a prospective study\cite{Lennon2020} compared with a median  across cancers of 70\% in a retrospective study.\cite{Cohen2018} Our model projects that a modest early-stage test sensitivity will translate into a similarly modest reduction in advanced-stage diagnoses. The implications for mortality will vary across cancers but, in general, the mortality reduction is expected to be lower than the reduction in advanced stage diagnoses.\cite{Owens2022}

Second, our framework for modeling outcomes under prospective screening permits quantifying overdiagnosis, or the detection by screening of cancers that would not be diagnosed without screening. Because this outcome is not observable and requires additional data on competing mortality that may differ in screened individuals from the general population, we did not attempt to estimate this outcome in this study. As early data from MCED testing becomes available, overdiagnosis is certain to be crucially important outcome to quantify.

Third, our framework can directly inform the design of MCED screening trials. Hu and Zelen\cite{Hu2004} used modeling for trial design, focusing on sample size determination for trials with an endpoint of disease-specific mortality. Our simulation framework supports design of trials using stage shift as an endpoint or as a surrogate endpoint for disease-specific mortality. For example, our model can be readily extended to predict mortality reductions based on the stage-shift multiplier of Owens et al.\cite{Owens2022} The modeling framework presented herein can be used to project outcomes for a wide range of trial designs to study the efficacy of an MCED test with given target cancers and cancer- and stage-specific sensitivities.

Our natural history model is necessarily a simplification of complex biological processes and as such is subject to several limitations. We require that disease stage be dichotomized into early versus advanced stage, but we allow the dichotomy to differ across cancers. We did not estimate distinct models for different histologic subtypes, although it would be possible to do so. We assumed transition rates between states are constant over time, which corresponds to exponential dwelling time distributions; more flexible distributions are not identifiable from the data inputs considered. We also assume constant test sensitivities within each stage; more flexible dynamics of test sensitivity are also not identifiable without from the data inputs considered. Finally, we did not account for uncertainty in the data inputs nor  stochastic uncertainty in representing finite populations, but these are feasible extensions.

Despite these limitations, our framework provides a transparent system for evidence synthesis and a mechanistic tool to translate from single-cancer and MCED screening test performance to a key driver of benefit.

\clearpage


\newpage
\begin{center}
\textbf{\large Supplemental Materials: A general framework combining individual cancer models to project
downstaging due to multi-cancer early detection testing}
\end{center}
\setcounter{section}{0}
\setcounter{equation}{0}
\setcounter{figure}{0}
\setcounter{table}{0}
\setcounter{page}{1}
\makeatletter

\renewcommand{\thesection}{S-\arabic{section}}
\renewcommand{\theequation}{S\arabic{equation}}
\renewcommand{\thefigure}{S\arabic{figure}}
\renewcommand{\thetable}{S\arabic{table}}

\section{The general two-stage progressive natural history model}

\subsection{Model states and transition rates}\label{sup:model}

\begin{figure}[h!]
    \centering
    \includegraphics[scale=0.8]{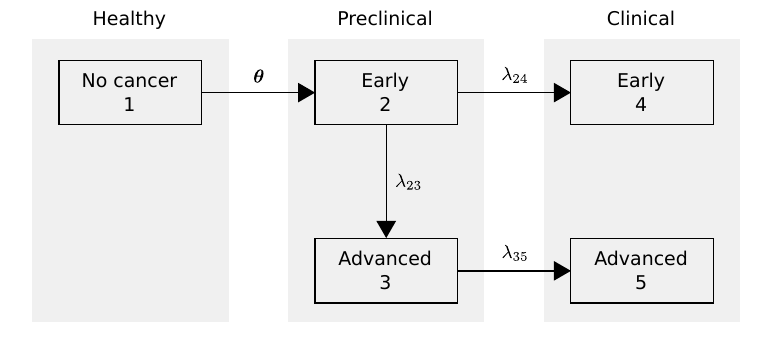}
    \caption{General two-stage progressive natural history model}
    \label{fig:natural history}
\end{figure}

The natural history model comprises a state space $S=\{$1 = no cancer, 2 = early-stage preclinical cancer, 3 = advanced-stage preclinical cancer, 4 = early-stage clinical cancer, 5 = advanced-stage clinical cancer$\}$ and parameters governing transition rates between states (Figure~\ref{fig:natural history}).

\begin{figure}[h!]
    \centering
    \includegraphics[scale=0.8]{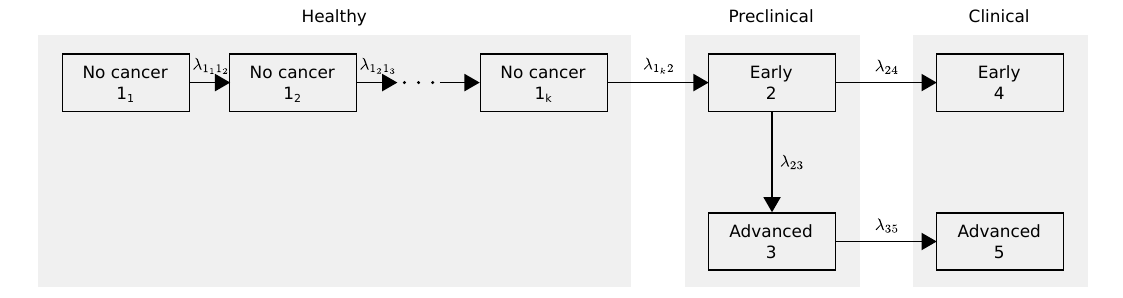}
    \caption{Two-stage progressive natural history model with hypoexponential preclinical onset distribution}
    \label{fig:CTMChistory}
\end{figure}

Let $g_1(u)$ be the age-dependent density of clinical onset and denote the constant rates of transitioning between states 2, 3, and 4 as $\lambda_{23}$, $\lambda_{24}$, and $\lambda_{35}$. We use a hypoexponential density for $g_1(u)$, which is equivalent to a series of exponentially distributed transitions governed by a set of parameters $\boldsymbol{\theta}=\{\lambda_{1_1 1_2},\lambda_{1_2 1_3},\ldots,\lambda_{1_k 2}\}$ (Figure~\ref{fig:CTMChistory}).

\subsection{Hazard functions for early- and advanced-stage clinical diagnoses}\label{sup:hazard}

The parameters that govern transitions between states can be used to calculate hazards of early- and advanced-stage clinical diagnosis with respect to age $a$, denoted $h_{4}(a)$ and $h_{5}(a)$, respectively. We calculate these cause-specific hazard functions by considering early- and advanced-stage clinical diagnoses as competing states. The probability of transitioning to state $k$ by age $a$ is represented as $Pr(A \leq a, K = k)$, where $A$ is the age at clinical diagnosis and $K \in \{4,5\}$ indicates the stage of clinical diagnosis. The probability of not transitioning to either state by age $a$ is represented by:
\[
   S(a) = 1 - \left[\Pr(A \leq a, K = 4) + \Pr(A \leq a, K = 5)\right].
\]

\noindent
The cause-specific density, or the unconditional rate of failures by age $a$, is calculated as:
\[
    f_k(a) = \lim_{dt\to0}\frac{\Pr\left\{A \in [a,a + dt), K = k\right\}}{dt}.
\]

\noindent
Finally, the cause-specific hazard, or the instantaneous probability of transitioning to state $k$ in those who have survived to age $a$, is calculated as:
\[
    h_k(a) = \frac{f_k(a)}{S(a)}
\]
for $k \in \{4,5\}$.

Suppose $u$ is the time of preclinical onset (i.e., the latent transition from state 1 to 2) and $w$ is the time of progression to advanced-stage disease (i.e., the latent transition from state 2 to 3). The cause-specific density of early-stage clinical diagnosis at age $a$ marginalized over $u$ is:
\[
    f_{4}(a)=\int_0^a g_1(u) \exp\left[-(\lambda_{23}+\lambda_{24})(a-u)\right]\lambda_{24}\; du.
\]

\noindent
Similarly, the cause-specific density of advanced-stage clinical diagnosis at age $a$ marginalized over $u$ and $w$ is:
\[
    f_{5}(a)=\int_0^a \int_u^a g_1(u) \exp\left[-(\lambda_{23}+\lambda_{24})(w-u)\right]\lambda_{23} \exp\left[-\lambda_{35}(a-w)\right]\lambda_{35}\; dw\; du.
\]

\noindent
Consequently, the probability of early-stage clinical diagnosis by age $a$ is given by:
\[
    \Pr(A \leq a, K = 4)=\int_0^a \int_0^s g_1(u) \exp\left[-(\lambda_{23}+\lambda_{24})(s-u)\right]\lambda_{24}\; du\; ds.
\]

\noindent
And the probability of advanced-stage clinical diagnosis by age $a$ is given by:
\[
    \Pr(A \leq a, K = 5)=\int_0^a \int_0^s \int_u^s g_1(u) \exp\left[-(\lambda_{23}+\lambda_{24})(w-u)\right)\lambda_{23} 
\exp\left(-\lambda_{35}(s-w)\right]\lambda_{35}
\;dw \; du \; ds.
\]

\subsection{Latent CTMC method for computing cause-specific hazards}\label{sup:latenthazard}

Although the calculation of cause-specific hazard functions can be done via numeric integration, it is computationally expensive. To overcome this, we employed a latent continuous time Markov chain (CTMC) approach. This approach uses the same hypoexponential distribution for preclinical onset and exponential distributions for transitions from early- to advanced-stage preclinical cancer and from preclinical to clinical states. The progression from healthy to clinical disease states is described by the multistate process $W(a)$, with state space $S$, and is based on an underlying CTMC $X(a)$, with state space G. The latent CTMC approach assumes that latent states in $S$ correspond to disease states in $G$, allowing for non-exponentially distributed sojourn times for $W(a)$. This approach allows for latent states in the preclinical onset distribution only (i.e., the transition from state 1 to 2), resulting in the latent state space $G = {1_1, \ldots, 1_k, 2, 3, 4,5}$, where $k$ is the number of latent preclinical states. The mapping of $S$ into $G$ is:
\[
\begin{aligned}
W(a)&=1 \longleftrightarrow X(a)\in \{1_1, \ldots, 1_k\} \\
W(a)&=2 \longleftrightarrow X(a)=2 \\
W(a)&=3 \longleftrightarrow X(a)=3 \\
W(a)&=4 \longleftrightarrow X(a)=4 \\
W(a)&=5 \longleftrightarrow X(a)=5.
\end{aligned}
\] 

The underlying CTMC $X(a)$ follows a time-homogeneous CTMC, which is parameterized via an intensity matrix $\boldsymbol \Lambda = \{\lambda_{ij}\}$, $i,j \in G$, $i<j$, that describes rates of transitions between latent states and an initial distribution $\boldsymbol{\pi}$ that assumes individuals start in state $1_1$ at age 0. The intensity matrix is given by:
\[
    \boldsymbol\Lambda = \left[
        \begin{array}{cccccccccc}
-\lambda_{1_1 1_2} & \lambda_{1_1 1_2} & 0 & \cdots & & & & & \cdots & 0 \\
0 &-\lambda_{1_2 1_3} & \lambda_{1_2 1_3} & \cdots & & & & & \cdots & 0 \\
\vdots & & \ddots & \ddots & & & & & & \vdots \\
0 & \cdots & & \cdots & -\lambda_{1_{k-1} 1_k} & \lambda_{1_{k-1} 1_{k}} & 0 & 0 & 0 & 0\\
0 & \cdots & & \cdots & 0 & -\lambda_{1_k 2} & \lambda_{1_k 2} & 0 & 0 & 0\\
0 & \cdots & & \cdots & 0 & 0 & -(\lambda_{23}+\lambda_{2 4}) & \lambda_{23} & \lambda_{24} & 0\\
0 & \cdots & & \cdots & 0 & 0 & 0 & -\lambda_{35} & 0 & \lambda_{35} \\
        \end{array} \right].
\] 

Suppose that $\mathbf{P}(t) = \{P_{ij}(t)\}$ is the matrix representing transition probabilities between states $i$ and $j$ in the interval $[s,s+t]$. The transition probability matrices between states are characterized by a matrix exponential of the intensity matrix $\boldsymbol \Lambda$:
\[
    \mathbf{P}(t)=\exp(\boldsymbol \Lambda t).
\]

\noindent
This characterization enables us to calculate probabilities of observing early- and advanced-stage clinical diagnosis by age $a$ as follows:
\[
    \Pr(A \leq a, K = 4)=P_{1_1 4}(a)
\] 
and
\[
    \Pr(A \leq a, K = 5)=P_{1_1 5}(a).
\]

\noindent
In general, the density of the time transition to state $j$ at time $s+t$, given that the process is in state $i$ at time $s$, is:
\[
    f_{ij}(t)=\sum_{\ell \in S} P_{i\ell}(t)\lambda_{\ell j}.
\]

\noindent
The cause-specific densities of early and advanced-stage clinical diagnosis at age $a$ can be expressed as:
\[
    f_4(a)=f_{1_1 4}(a)
\]
and
\[
    f_5(a)=f_{1_1 5}(a).
\]

\noindent
Thus, the cause-specific hazard for early-stage clinical diagnosis at age $a$ is:
\[
    h_4(a)=\frac{f_{1_1 4}(a)}{1-\left[P_{1_1 4}(a)+P_{1_1 5}(a)\right]}.
\]

\noindent
Similarly, cause-specific hazard for advanced-stage clinical cancer at age $a$ is:
\[
    h_5(a)=\frac{f_{1_1 5}(a)}{1-\left[P_{1_1 4}(a)+P_{1_1 5}(a)\right]}.
\]

\subsection{Model parameterization and identifiability}\label{sup:identifiability}

The full set of parameters governing the natural history $\{\lambda_{ij}\}$, $i,j \in G$, $i<j$ are not fully identifiable given only the incidence data. The parameters $\lambda_{23},\lambda_{24},\lambda_{35}$ and the onset density $g_1(u)$
can be identified given hypothesized values for the overall mean sojourn time (OMST) $M$ and the advanced-stage mean sojourn time (LMST) $\frac{1}{\lambda_{35}}$. The OMST can be written:
\[
    \text{OMST}=\frac{1}{\lambda_{24}+\lambda_{23}}+\frac{\lambda_{23}}{\lambda_{23}+\lambda_{24}}\frac{1}{\lambda_{35}}.\\
\]

\noindent
In practice, we estimate $\lambda_{23}$ and express $\lambda_{24}$ in terms of the OMST, $\lambda_{23}$, and $\lambda_{35}$:
\[
    \lambda_{24}=\frac{\lambda_{35}+\lambda_{23}}{\lambda_{35}\times \text{OMST}}-\lambda_{23}.\\
\]

\noindent
This expression also helps us bound $\lambda_{23}$. Because $\lambda_{23},\lambda_{24},\lambda_{35}>0$, we know that $0<\lambda_{23}<\frac{\lambda_{35}}{\lambda_{35}\times \text{OMST}-1}$. This also implies we can only consider $\text{LMST}<\text{OMST}$.

We note that even with these hypothesized values for the OMST and LMST, the parameters in $\boldsymbol{\theta}$ governing the hypoexponential onset distribution are not guaranteed to be identifiable, which is a general characteristic of latent Markov models. Nonetheless, key functions of these parameters, such as the density, hazard, and cumulative distribution function, are identifiable (see, e.g., Bladt et al. \cite{Bladt2003}). In practice, for the settings considered in this paper, the non-uniqueness of $\boldsymbol{\theta}$ did not affect the identifiability of $\lambda_{23},\lambda_{24},\lambda_{35}$ given hypothesized values for the OMST and LMST.

In general, we do not specify in advance the choice of dimension $k$ for the onset distribution, as different cancer sites may be optimally fitted with different values of $k$. In practice, the dimension $k$ is specified by the user, typically based on visual inspection, to represent the minimal dimension that confers adequate flexibility. That is, we recommend an iterative selection approach, starting with $k=1$ and increasing $k$ until there is minimal incremental improvement in the deviation between observed and projected age- and stage-specific incidence.

\subsection{Fitting the model using a Poisson likelihood}\label{sup:likelihood}

We obtained early- and advanced-stage incidence counts from the core 9 registries of the Surveillance, Epidemiology, and End Results (SEER) program tabulated by age in 5-year groups: 0-4, 5–9, \ldots, 80–84, and 85+ years. We assumed that the number of early-stage ($o_{ae}$) and advanced-stage ($o_{a\ell}$) diagnoses in each cell followed a Poisson distribution with mean:
\[
    M_{ak}=\text{PY}(a) \times h_k(a),
\]
where $k\in \left\{4,5\right\}$, $a$ is the midpoint age of the interval, $\text{PY}(a)$ is the number of person-years at risk during the interval containing $a$ (i.e., the population count $\times$ the length of the interval). Then the overall likelihood is given by:
\[
    \mathcal{L}=\prod_{a,k} \frac{M_{ak} \exp(-M_{ak})}{o_{ak}}.
\]

\noindent
We fit the exogenous model parameters by maximizing this likelihood using a numeric optimization algorithm\cite{Liu1989} with box constraints. 

\section{Projecting stage shift from a fitted model}\label{sup:stageshift}

To project stage shift due to a screening test, we used the estimated model to simulate advanced-stage diagnosis with (intervention arm) and without (control arm) screening in a hypothetical trial. The trial only enrolls individuals with no prior clinical diagnosis at the start of screening. The trial design includes screens at ages $a_1, \ldots, a_n$ and final follow-up at age $a_{n+1}$. Test sensitivities for early- and advanced-stage preclinical cancers are represented by $B_e$ and $B_\ell$, respectively. The stage shift represents the relative reduction in advanced-stage diagnoses between the intervention and control arms at the end of the trial.

\subsection{Interval-specific stage shift}

In the intervention arm, let $S_D(a_\ell)$  be the probability of screen-detected advanced-stage cancer at age $a_\ell, \ell\in1,\ldots,n$ and let $J_D(a_\ell, a_{\ell+1})$ be the probability of advanced-stage clinical diagnoses in the interval $(a_\ell,a_{\ell+1}), \ell\in1,\ldots,n$. In the control arm, let $C_D[a_\ell,a_{\ell+1})$ be the probability of advanced-stage clinical diagnoses in the interval $[a_\ell,a_{\ell+1}), \ell\in1,\ldots,n$. Then, for a given interval ($\ell\in1,\ldots,n)$, the projected interval-specific stage shift is:
\[
    \frac{C_D[a_\ell,a_{\ell+1})-\left\{S_D(a_\ell)+J_D(a_\ell, a_{\ell+1})\right\}}{C_D[a_\ell,a_{\ell+1})}.
\]

\subsection{Cumulative stage shift}

The cumulative stage shift at the end of the trial is:
\[
    \frac{\sum_{\ell=1}^{n+k-1}C_D[a_\ell,a_{\ell+1})-\left\{\sum_{\ell=1}^nS_D(a_\ell)+\sum_{\ell=1}^{n}J_{D}(t_\ell,t_{\ell+1})\right\}}{\sum_{\ell=1}^{n}C_D[a_\ell,t_{\ell+1})}.
\]

\subsection{Computing the stage shift probabilities}
\subsubsection{Control arm}

In the control arm, the probability of advanced-stage diagnosis between ages $a_\ell$ and $a_{\ell+1}$ is:
\[
    C_D[a_\ell,a_{\ell+1})=\frac{P_{{1_1}5}(a_{\ell+1})-P_{{1_1}5}(a_\ell)}{1-\left[P_{{1_1}4}(a_1)+P_{{1_1}5}(a_1)\right]},
\]
where the denominator indicates the probability of no clinical diagnosis before the start of the trial.

\subsubsection{Intervention arm}

To compute the relevant quantities in the intervention arm, we denote observed states at each screen or follow-up time as $O_1,\ldots,O_{n+1} \in S=\{$1 = no cancer, 2 = early-stage preclinical cancer, 3 = advanced-stage preclinical cancer, 4 = early-stage clinical cancer, 5 = advanced-stage clinical cancer$\}.$ With this notation, the probability of screen-detected advanced-stage cancer at the first screen at age $a_1$ is:
\[
    S_D(a_1)=P(O_1=3),
\]
and generally for subsequent screens at a age $a_\ell$, $\ell\in2,\dots,n$, is:
\[
    S_D(a_\ell)=P(O_1=1,\ldots,O_{\ell-1}=1,O_\ell=3).
\]

\noindent
Likewise, the probability of clinically diagnosed cancer in the interval $(a_\ell,a_{\ell+1})$, $\ell\in1,\dots,n$, is:
\[
    J_D(a_\ell,a_{\ell+1})=P(O_1=1,\ldots,O_{\ell}=1,O_{\ell+1}=5).
\]

To obtain these quantities, suppose that $x_1, \ldots, x_\ell$ are the states of the underlying disease process at ages $a_1, \ldots, a_\ell$. The probability of the observed data is constructed by determining the joint probability of $x_1, \ldots, x_\ell$ and $o_1, \ldots, o_\ell$, and then marginalizing across $x_1, \ldots, x_\ell$: 
\[
\begin{aligned}
\Pr(O_1=o_1, \ldots, O_\ell=o_\ell)=\sum_{x_1 \in S} \cdots \sum_{x_\ell \in S}
& \Pr(O_1=o_1, \ldots, O_\ell=o_\ell\mid X_1=x_1, \ldots, X_\ell=x_\ell)\\
& \times \Pr(X_1=x_1, \ldots, X_\ell=x_\ell).
\end{aligned}
\]

\noindent
This probability can be simplified based on the Markov property of $X(a)$, where the underlying state at age $a_k$ only depends on prior observations through age $a_{k-1}$. We also assume that the observation $O_{i}$ given $X_{i}$ is conditionally independent of the underlying disease states at other times. The probability of observing a particular sequence of outcomes is given by:

\begin{equation}\label{eq:prob}
    \Pr(O_1=o_1, \ldots, O_\ell=o_\ell)=\sum_{x_1\in S}\cdots\sum_{x_\ell \in S} \pi_{x_1}\prod_{i=1}^{\ell-1} P_{x_i,x_{i+1}}(a_{i+1}-a_i) \prod_{i=1}^\ell P(O_i=o_i\mid X_i=x_i).
\end{equation}

\noindent
Here the probability of observing a value of $O_i$ given the underlying value $X_i$ is described by the $(k+4)\times 5$ matrix $\left\{E[r,s]\right\}$ with entries $E[r,s]=P(O_i=s\mid X_i=r)$:
\[
\mathbf{\text{E}}=
\begin{bmatrix}
1 & 0 & 0 & 0 & 0 \\
1 & 0 & 0 & 0 & 0 \\
\vdots & \vdots & \vdots & \vdots & \vdots \\
1 & 0 & 0 & 0 & 0 \\
1-B_e & B_e & 0 & 0 & 0 \\    
1-B_\ell & 0 &  B_\ell & 0 & 0 \\  
0 & 0 & 0 & 1 & 0 \\
0 & 0 & 0 & 0 & 1\\
\end{bmatrix}. 
\]

\noindent
The likelihood also denotes the initial distribution of $X(a_1)$ at age $a_1$, $\boldsymbol{\pi}$, where a criterion for entering the trial is not having had clinical diagnosis prior to $a_1$, resulting in $\pi_j=0$ for $j\in{4,5}$. For $j\in{1_1,\ldots,1_k,2,3}$:
\[
    \pi_j=\frac{P_{1_1j}(a_1)}{1-\left[P_{1_1,4}(a_1)+P_{1_1,5}(a_1)\right]}.
\]

\noindent
The formulation in Eq. \ref{eq:prob} is computationally equivalent to the likelihood for a hidden Markov model and can be computed efficiently using the Baum-Welch forward-backward algorithm.\cite{Baum1970}

\section{Converting the MISCAN-Lung model parameters to exogeneous parameters in our lung cancer model}\label{sup:weightedparam}

\subsection{Sensitivities of low-dose computed tomography and x-ray radiography}

The MISCAN-Lung model specifies sensitivities by the American Joint Committee on Cancer 7th Edition (AJCC7) by stage and histology. Here we derived estimates for early- and advanced-stage sensitivities for low dose computed tomograph (LDCT) and x-ray radiography across histologies, where early-stage is defined as AJCC7 I--II and advanced stage is defined as AJCC7 III--IV. Based on the stage- and histology-specific proportions of diagnoses in the National Lung Screening Trial (NLST) population\cite{Team2011} (Supplement Table~\ref{SuppTab:propstagehist}), we computed weighted averages of sensitivities across AJCC7 stages and histologies (Supplement Table~\ref{SuppTab:sens}). 

\begin{table}
\centering
\caption{Proportions of diagnoses by stage and histology in the National Lung Screening Trial}
\label{SuppTab:propstagehist}
\begin{tabular}{cccccc}
\hline
Aggregate & AJCC7 & Adenocarcinoma & Squamous cell & Small cell & Other \\
stage & stage & (35\%) & carcinoma (22\%) & lung cancer (17\%) &(26\%)  \\ 
\hline
Early    & 1A   & 0.66 & 0.66 & 0.63 & 0.64 \\
         & 1B   & 0.20 & 0.20 & 0.22 & 0.21 \\
         & IIB  & 0.15 & 0.14 & 0.15 & 0.15 \\
\hline
Advanced & IIIA & 0.33 & 0.34 & 0.33 & 0.33 \\
         & IIIB & 0.24 & 0.24 & 0.24 & 0.23 \\
         & IV   & 0.43 & 0.41 & 0.43 & 0.44 \\
\hline
\end{tabular}
\end{table}

\begin{table}
\centering
\caption{Weighted means of sensitivity of low-dose computed tomography and x-ray radiography in the MISCAN-Lung model by stage and histology}
\label{SuppTab:sens}
\begin{tabular}{cccccc}
\hline
AJCC7 & Adenocarcinoma & Squamous cell & Small cell & Other & Weighted mean \\
stage & & carcinoma & lung cancer & & by histology \\
\hline
\multicolumn{6}{l}{\textbf{LDCT sensitivity for early-stage  lung cancer}} \\
1A   & 56.6 & 31.0 & 8.83 & 20.8 & 33.5 \\
1B   & 64.1 & 38.1 & 10.3 & 24.8 & 39.0 \\
IIB  & 64.5 & 39.2 & 11.2 & 24.8 & 39.5 \\
Weighted mean by stage & 59.3 & 33.6 & 9.51 & 22.2 & \textbf{35.5} \\ 
\hline
\multicolumn{6}{l}{\textbf{LDCT sensitivity for advanced-stage lung cancer}} \\
IIIA & 75.9 & 69.7 & 41.6 & 60.4 & 64.7 \\
IIIB & 80.2 & 79.4 & 87.1 & 68.3 & 78.1 \\
IV   & 98.9 & 97.7 & 99.4 & 95.7 & 97.9 \\
Weighted mean by stage & 86.9 & 83.6 & 77.3 & 77.8 & \textbf{82.1} \\ 
\hline
\multicolumn{6}{l}{\textbf{X-ray radiography sensitivity for early-stage lung cancer}} \\
1A   & 16.9 & 9.70 & 2.51 & 6.27 & 10.1 \\
1B   & 27.1 & 28.9 & 4.25 & 7.57 & 18.5 \\
IIB  & 27.3 & 30.0 & 6.64 & 7.57 & 19.2 \\
Weighted mean by stage & 20.4 & 16.5 & 3.52 & 6.74 & \textbf{13.1} \\
\hline
\multicolumn{6}{l}{\textbf{X-ray radiography sensitivity for advanced-stage lung cancer}} \\
IIIA & 48.1 & 46.3 & 14.7 & 29.8 & 37.3 \\
IIIB & 49.3 & 48.0 & 53.2 & 34.4 & 45.8 \\
IV   & 96.3 & 78.6 & 97.3 & 36.9 & 77.2 \\
Weighted mean by stage & 69.2 & 60.0 & 59.5 & 34.0 & \textbf{56.4} \\
\hline
\end{tabular}
\end{table}

\subsection{Overall and advanced-stage mean sojourn times}

The MISCAN-Lung model estimated mean sojourn times by AJCC7 stage, sex, and histology and probabilities of transitioning between stages by histology. Here we derived estimates of the overall mean sojourn time (OMST) and the advanced-stage mean sojourn time (LMST) from the MISCAN-Lung model:
\begin{enumerate}
\item The overall mean sojourn time (OMST), which represents the overall time spent in any stage, conditional on starting in state IA.
\item The early mean sojourn time (EMST), which represents the overall time spent in stages I--II, conditional on starting in state IA.
\item The late mean sojourn time (LMST), which represents the overall time spent in stages III--IV, conditional on entering stage III.
\end{enumerate}

\noindent
For each histology, we first computed the weighted mean sojourn time $\text{MST}(x)$ for each stage $x$ across sexes based on proportions in the NLST (Supplement Table~\ref{SuppTab:sojurn}). The OMST is calculated by summing the mean duration spent in each stage, conditional on starting in stage IA. For stage $x$, the mean duration spent in stage $x$ is $\Pr(\text{IA to }x)\times \text{MST}(x)$, where $\Pr(\text{IA to }x)$ is the probability of being in stage $x$ conditional on starting in state $IA$ and is derived from the transition probabilities between successive stages in the MISCAN-Lung model (Supplement Table~\ref{SuppTab:transprob}). For example:
\[
    \Pr(\text{IA to II})=\Pr(\text{IA to IB})\times \Pr(\text{IB to II}).
\]

\noindent
Thus, for a given histology, the OMST is:
\[
\begin{aligned}
    \text{OMST}=&\text{MST}(\text{IA}) \\
                &+\Pr(\text{IA to IB})\times \text{MST}(\text{IB}) \\
                &+\Pr(\text{IA to II})\times \text{MST}(\text{II}) \\
                &+\Pr(\text{IA to III})\times\text{MST}(\text{III}) \\
                &+\Pr(\text{IA to IV})\times \text{MST}(\text{IV}).
\end{aligned}
\]

\noindent
Similarly, for a given histology, the EMST is the mean duration of time spent in stages I--II, conditional on starting in state IA, and is given by:
\[
\begin{aligned}
    \text{EMST}=&\text{MST}(\text{IA}) \\
                &+\Pr(\text{IA to IB})\times \text{MST}(\text{IB}) \\
                &+\Pr(\text{IA to II})\times \text{MST}(\text{II}).
\end{aligned}
\]

\noindent
For a given histology, the LMST is the mean duration of time spent in stage III--IV, conditional on entering stage III, and is given by:
\[
\begin{aligned}
    \text{LMST}=&\text{MST}(\text{IIIA}) \\
                &+\Pr(\text{IIIA to IIIB})\times \text{MST}(\text{IIIB}) \\
                &+\Pr(\text{IIIA to IV})\times \text{MST}(\text{IV}).
\end{aligned}
\]

\noindent
Supplement Table~\ref{SuppTab:OMST} provides the calculated OMST, EMST, and LMST by histologic type and the weighted average across types, which we use as inputs in our lung cancer model.

\begin{table}
\centering
\caption{Mean sojourn by stage times estimated in the MISCAN-Lung model averaged across men (59\%) and women (41\%)}
\label{SuppTab:sojurn}
\begin{tabular}{ccccc}
\hline
Stage & Adenocarcinoma & Squamous cell carcinoma & Small cell lung cancer & Other \\
\hline
IA   & 2.07 & 2.16 & 1.30 & 2.10 \\
IB   & 0.73 & 0.76 & 0.46 & 0.74 \\
II   & 0.53 & 0.55 & 0.33 & 0.54 \\
IIIA & 0.53 & 0.55 & 0.33 & 0.54 \\
IIIB & 0.41 & 0.42 & 0.26 & 0.41 \\
IV   & 0.84 & 0.88 & 0.53 & 0.86 \\ 
\hline
\end{tabular}
\end{table}
  
\begin{table}
\centering
\caption{Probabilities of transitioning between preclinical stages in the MISCAN-Lung model}
\label{SuppTab:transprob}
\begin{tabular}{ccccc}
\hline
Transition & Adenocarcinoma & Squamous cell carcinoma & Small cell lung cancer & Other \\
\hline
IA to IB     & 0.85 & 0.87 & 0.97 & 0.92  \\
IB to II     & 0.88 & 0.85 & 0.97 & 0.94  \\
II to IIIA   & 0.93 & 0.87 & 0.97 & 0.95  \\
IIIA to IIIB & 0.87 & 0.81 & 0.89 & 0.87  \\
IIIB to IV   & 0.76 & 0.65 & 0.80  & 0.80 \\
\hline
\end{tabular}
\end{table}

\begin{table}
\centering
\caption{Overall, early-, and advanced-stage mean sojourn times by histology averaged across sexes}
\label{SuppTab:OMST}
\begin{tabular}{cccccc}
\hline
Sojourn time & Adenocarcinoma & Squamous cell carcinoma & Small cell lung cancer & Other & Weighted mean \\ 
\hline
Overall        & 4.09 & 4.09 & 2.90 & 4.48 & \textbf{3.99} \\
Early stage    & 3.09 & 3.22 & 2.05 & 3.25 & \textbf{2.98} \\
Advanced stage & 1.44 & 1.35 & 0.94 & 1.49 & \textbf{1.35} \\
\hline
\end{tabular}
\end{table}

\clearpage
\bibliographystyle{naturemag}
\bibliography{stageshiftbib.bib}
\end{document}